\documentclass[12pt]{iopart}
\usepackage{latexsym}
\usepackage{amscd}
\usepackage[dvips]{color}

\begin{document}
\jl{31} 
 \def\lambdabar{\protect\@lambdabar}
\def\@lambdabar{%
\relax
\bgroup
\def\@tempa{\hbox{\raise.73\ht0
\hbox to0pt{\kern.25\wd0\vrule width.5\wd0
height.1pt depth.1pt\hss}\box0}}%
\mathchoice{\setbox0\hbox{$\displaystyle\lambda$}\@tempa}%
{\setbox0\hbox{$\textstyle\lambda$}\@tempa}%
{\setbox0\hbox{$\scriptstyle\lambda$}\@tempa}%
{\setbox0\hbox{$\scriptscriptstyle\lambda$}\@tempa}%
\egroup}
\definecolor{r}{rgb}{1,0,0}

\def\bbox#1{%
\relax\ifmmode
\mathchoice
{{\hbox{\boldmath$\displaystyle#1$}}}%
{{\hbox{\boldmath$\textstyle#1$}}}%
{{\hbox{\boldmath$\scriptstyle#1$}}}%
{{\hbox{\boldmath$\scriptscriptstyle#1$}}}%
\else
\mbox{#1}%
\fi
}
\newcommand{\beq}{\begin{equation}}
\newcommand{\eeq}{\end{equation}}
\newcommand{\beqan}{\begin{eqnarray*}}
\newcommand{\eeqan}{\end{eqnarray*}}
\newcommand{\beqa}{\begin{eqnarray}}
\newcommand{\eeqa}{\end{eqnarray}}
\newcommand{\hX}{\widehat{\bbox{X}}}
\newcommand{\hM}{\widehat{\bbox{M}}}
\newcommand{\hL}{\widehat{\bbox{L}}}
\newcommand{\hT}{\widehat{\bbox{T}}}
\newcommand{\hQ}{\widehat{\bbox{Q}}}
\newcommand{\eps}{\varepsilon}
\newcommand{\la}{\lambda}
\newcommand{\scr}{\scriptstyle}
\newcommand{\al}{\alpha}
\renewcommand{\d}{\partial}
\def\rmi{{\rm i}}
\def\rme{\bbox{\rm e}}
\def\rmd{\bbox{\rm d}}
\newcommand{\Bv}{\bbox{B}}
\newcommand{\Ev}{\bbox{E}}
\newcommand{\Av}{\bbox{A}}
\newcommand{\jv}{\bbox{j}}
\newcommand{\rv}{\bbox{r}}
\newcommand{\dom}{\mathcal D}
\newcommand{\intl}{\int\limits}
\newcommand{\suml}{\sum\limits}
\newcommand{\nablav}{\bbox{\nabla}}
\newcommand{\teps}{\tilde{\eps_0}}
\newcommand{\tmu}{\tilde{\mu_0}}
 \newcommand{\Mv}{\bbox{M}}
\newcommand{\Pv}{\bbox{P}}
\newcommand{\Rv}{\bbox{R}}
\newcommand{\pv}{\bbox{p}}
\newcommand{\mv}{\bbox{m}}
\newcommand{\Pcal}{\mathcal P}
\newcommand{\Mcal}{\mathcal M}
\newcommand{\Dv}{\bbox{D}}
\newcommand{\Hv}{\bbox{H}}
\newcommand{\gam}{\gamma}
\newcommand{\ev}{\bbox{e}}
\title{ On the  formulation of Electrodynamics in a form independent of units}
\author{C. Vrejoiu  }
\address{Department of Physics, University of Bucharest, PO Box MG-11,
 Bucharest, Romania\\  E-mail : cvrejoiu@fizica.unibuc.ro  }
\begin{abstract}
Some pedagogical aspects of the formulation of electrodynamics independently of the  systems of units are commented. Only electrodynamics in vacuum is considered. It is  pointed out the efficiency of using a notation system close to  {\bf SI}.
\end{abstract}
\section{Introduction}
Certainly, the simplest system  of units  is {\bf SI}. 
  Most of the students know and use only {\bf SI}, as science and technology notions are mainly taught only in this system of units. This is not necessarily a bad aspect. However, there are many students, especially above average ones, who demand additional knowledge to the one presented in regular faculty lectures. For them, other systems of units are necessary - at least Gauss and Heaviside-Lorentz if we refer to  electrodynamics. This motivates the interest on teaching electrodynamics in a notation scheme independent of the system of units.  
  One of the latest issues on this subject is Ref. \cite{Heras} where one can find a detailed bibliography.
\section{ ``System free'' expressions of Maxwell's equations}
\par  Usually, in an introductory course in the electromagnetic theory  one introduces, as a first step, the basic variables $\Ev,\,\Bv$ of the electromagnetic field  by postulating the expression of the Lorentz force as an experimental equation:
\beqa\label{LF}
\bbox{F}\,=\,q\,\Ev\,+\,\tilde{\gam}\,q\,\bbox{v}\times\Bv\ 
\eeqa
where $\tilde{\gam}$ is a proportional constant introduced so as to account for different dimensions and units of measurements of electrodynamic quantities. This expression is sufficient for introducing the electric charge, the electric field $ \Ev$ by the force acting on a charged particle at rest and the magnetic field $\Bv$ as responsible for the part of Lorentz force due to the particle motion. 
The electric charge conservation is postulated  by introducing the  continuity equation,
\beqa\label{ecc}
\frac{\d\rho}{\d t}\,+\,\nablav\cdot\jv\,=\,0\ ,
\eeqa
 a necessary  equation for understanding the  Maxwell's postulate of the displacement current .
\par It is possible to adopt a different treatment, specific for a general field theory: the fields $\Ev$ and $\Bv$ are introduced as basic variables verifying the Maxwell's equations. It remains to clarify the role of these variables in the description of the interactions between the electromagnetic field and other physical systems together with the formulation of general theorems as electric charge, energy, linear  and angular momentum theorems.
\footnote{ However, using variational and symmetry  principles should be the better way.}
\par 

\par Let us  write Maxwell's equations in vacuum with notation used in \cite{Heras}:
\beqa\label{Maxwell-H}
(a)\;\;\;\;\; \nablav\cdot \Ev\,=\,\al\,\rho\ ,\nonumber\\
(b)\;\;\;\;\; \nablav\cdot\Bv\,=\,0\ ,\nonumber\\
(c)\;\;\;\;\; \nablav\times\Bv\,=\,\beta\,\jv\,+\,\frac{\beta}{\al}\,\frac{\d\Ev}{\d t}\ ,\nonumber\\
(d)\;\;\;\;\; \nablav\times\Ev\,=\,-\,\gam\,\frac{\d\Bv}{\d t}\ ,
\eeqa
with the proportional constants $\al,\,\beta,\,\gam$ depending only of units. The consequence of Maxwell equations concerning the propagation in vacuum of the electromagnetic field leads to the relation
\beqa\label{c}
\frac{\al}{\beta\gam}\,=\,c^2\ ,
\eeqa
$c$ being the speed of light in vacuum.  

\par The continuity equation for electric charge is a theorem resulting from Maxwell's equations \eref{Maxwell-H}-a, c. 
Together with this theorem, the definitions of electric charge density $\rho$ and the corresponding current density $\jv$ are accomplished. 
 \par Let us consider other two consequences of Maxwell's equations.   The first is the energy theorem 
 \beqa\label{u-th}
 \frac{\d u}{\d t}\,=\,-{\mathcal P}\,-\,\nablav\cdot\bbox{S}
 \eeqa
 where the expressions for the energy density $u$, energy density current $\bbox{S}$ and the spatial distribution of the power transfer between the field and particles must be introduced as basic postulates of the theory. 
 \par The linear momentum theorem is  represented by an equation obtained also from Maxwell's equations in the form:
 \beqa\label{m-th}
 \frac{\d\bbox{g}}{\d t}\,=\,-\bbox{f}\,-\,\ev_i\,\d_jT_{ji}\ ,
 \eeqa
 where the last expression from the right hand side represents the divergence $\nablav\cdot\bbox{\sf{T}}$ of a second order tensor and $\ev_i$ are the unit vectors of the Cartesian axes. As in the case of the energy theorem \eref{u-th}, in equation \eref{m-th} we have to identify the force density $\bbox{f}$, the momentum density $\bbox{g}$ and the momentum   current density $\bbox{\sf{T}}$, and postulate their expressions.

\par Just these definitions accompanying the formulation of the two theorems \eref{u-th} and \eref{m-th} represent the final step in the introduction of proportional constants $\al,\,\beta,\,\gam$ in the electromagnetic theory.

\par Let us introduce the set of equations resulting from Maxwell's equations whose combinations candidate for representing the two theorems \eref{u-th} and \eref{m-th}. From equations \eref{Maxwell-H}-c, d one obtains by a usual procedure:
\beqa\label{u1}
\frac{\d}{\d t}\left(\frac{\beta}{2\al}\,E^2\,+\,\frac{\gam}{2}\,B^2\right)\,=\,-\,\beta\,\jv\cdot\Ev\,-\,\nablav\cdot(\Ev\times\Bv)
\eeqa
which represents the candidate to the theorem of energy.
\par Let us the identity
\beqa\label{id}
\bbox{A}\times(\nablav\times\bbox{A})\,=\,(\nablav\cdot \bbox{A})\,\bbox{A}\, +\, \ev_i\,\d_j(\frac{1}{2}A^2\,\delta_{ji}\,-\,A_jA_i)\ ,
\eeqa
for an arbitrary vector $\bbox{A}$. Applying this identity to the fields $\Ev$ and $\Bv$ and using the first two equations \eref{Maxwell-H} one gets
\beqa\label{me}
\Ev\times(\nablav\times\Ev)\,=\,\al\,\rho\Ev\,+\,\ev_i\d_j(\frac{1}{2}E^2\,\delta_{ji}\,-\,E_jE_i)\ ,
\eeqa

and
\beqa\label{mb}
\Bv\times(\nablav\times\Bv)\,=\,\ev_i\d_j(\frac{1}{2}B^2\,\delta_{ji}\,-\,B_jB_i)\ .
\eeqa

Using Maxwell's equations \eref{Maxwell-H}-c, d, we obtain the two equations whose combination candidates to the momentum theorem:
\beqa\label{me1}
\gam\,\Ev\times\frac{\d\Bv}{\d t}\,=\,-\al\,\rho\Ev\,-\,\ev_i\d_j(\frac{1}{2}E^2\,\delta_{ji}\,-\,E_jE_i)\ ,
\eeqa

and
\beqa\label{mb1}
\frac{\beta}{\al}\frac{\d\Ev}{\d t}\times\Bv\,=\,-\beta\,\jv\times\Bv\,-\,\ev_i\d_j(\frac{1}{2}B^2\,\delta_{ji}\,-\,B_jB_i)\ .
\eeqa
In equations \eref{u1}, \eref{me1} and \eref{mb1}, the terms which represent the interaction between electromagnetic field and particles are only $\beta\,\jv\cdot\Ev,\;\;\al\,\rho\Ev\;$  and $\;\beta\,\jv\times\Bv$. A first choice of the place of proportional constants in different definitions is determined by choosing the units of force, electric charge and electric field such that, as is the case in all usual systems of units, the electric force density is $\bbox{f}_e\,=\,\rho\,\Ev$. For a correct appearance of the force density in the momentum theorem, the equation \eref{me1} must be considered divided by the constant $\al$. From this definition of $\bbox{f}_e$, the corresponding spatial distribution of the power transfer results to be ${\mathcal P}\,=\,\jv\cdot\Ev$. From this definition of  
${\mathcal P}$, and for the correct appearance in the energy theorem, equation \eref{u1} must be considered divided by $\beta$. 
The sum of the left sides of equations \eref{me1}, divided by $\al$,  and equation \eref{mb1}, appropriately modified, 
must represent a time derivative. For this end, the equation \eref{mb1} must be divided by $\beta$ and multiplied by $\gam$. 
\footnote{If we consider the alternative of using the expression \eref{LF} postulated before the introduction of Maxwell equations, this last operation consists in the multiplication of  equation \eref{mb1} by $\tilde{\gam}$ instead of $\gam$. In this case, the condition  that the sum of the right-hand sides of the two equations \eref{me1} and \eref{mb1}, after the accomplishment of these operations, represent a time derivative leads to the equality $\tilde{\gam}=\gam$, \cite{cv-04}. }

\par With these operations accomplished, in the equations derived from \eref{u1}, \eref{me1} and \eref{mb1} the constant $\beta$ is present only by the fraction
\beqa\label{la}
\la\,=\,\frac{\gam}{\beta}\ .
\eeqa
Using the constant $\la$ instead of $\beta$, these equations can be written as
\beqa\label{u11}
\frac{\d}{\d t}\left(\frac{1}{2\al}E^2\,+\,\frac{\la}{2}B^2\right)\,=\,-\,\jv\cdot\Ev\,-\,\nablav\cdot\left(\frac{\la}{\gam}\,\Ev\times\Bv\right)\ ,
\eeqa
\beqa\label{me11}
\frac{\gam}{\al}\,\Ev\times\frac{\d\Bv}{\d t}\,=\,-\rho\Ev\,-\,\ev_i\d_j\left(\frac{1}{2\al}E^2\,\delta_{ji}\,-\,\frac{1}{\al}E_jE_j\right)\ ,
\eeqa
and
\beqa\label{mb11}
\frac{\gam}{\al}\,\frac{\d\Ev}{\d t}\times\Bv\,=\,-\gam\,\jv\times\Bv\,-\,\ev_i\d_j\left(\frac{\la}{2}B^2\,\delta_{ji}\,-\,\la\,B_jB_i\right)\ .
\eeqa
Equation \eref{u11} represent the energy theorem \eref{u-th} with the definitions
\beqa\label{defu}
u\,=\,\frac{1}{2}\left(\frac{1}{\al}\,E^2\,+\,\la\,B^2\right),\;\;{\mathcal P}\,=\,\jv\cdot\Ev,\;\;\bbox{S}\,=\, \frac{\la}{\gam}\,\Ev\times\Bv\ .
\eeqa
The sum of equations \eref{me11} and \eref{mb11} represents the momentum theorem \eref{m-th} with the definitions
\beqa\label{defm}
\!\!\!\!\!\!\!\! \bbox{g}\,=\,\frac{\gam}{\al}\,\Ev\times\Bv,\;T_{ij}\,=\,u\,\delta_{ij}\,-\,\frac{1}{\al}\,E_iE_j\,-\,\la\,B_iB_j,\;
\bbox{f}\,=\,\rho\Ev\,+\,\gam\,\jv\times\Bv\ .
\eeqa
\par Equation \eref{c} is written as
\beqa\label{cc}
\frac{\al\la}{\gam^2}\,=\,c^2\ .
\eeqa
Let us write Maxwell's equations with the constants $\al,\,\la,\,\gam$:
\beqa\label{Maxwell-la}
a.\;\;\; \nablav\cdot \Ev\,=\,\al\,\rho\ ,\nonumber\\
b.\;\;\; \nablav\cdot\Bv\,=\,0\ ,\nonumber\\
c.\;\;\; \nablav\times\Bv\,=\,\frac{\gam}{\la}\big(\jv\,+\,\frac{1}{\al}\,\frac{\d\Ev}{\d t}\big)\ ,\nonumber\\
d.\;\;\; \nablav\times\Ev\,=\,-\,\gam\,\frac{\d\Bv}{\d t}\ ,
\eeqa

\par In Table 1 these constants are specialized, in a similar form as in Table 1 from \cite{Heras},   to the most used systems of units: Gauss and Heaviside-Lorentz  
\par 
\vspace{0.5cm}
\begin{center}
\par \scriptsize{Table 1. The $\al,\,\la,\,\gam$-system
\par
\begin{tabular}{@{}l@{}l @{}l}
\hline
System &~~~~~~~  $\al~~~~~$ &~ $\la$ ~~~~~~~~~~$\gam$\\
\hline
Gauss &~~~~~~~ $4\pi$~~~~~~~~&$1/4\pi$~~~~~~~ $1/c$ \\
SI~~~ &~~~~~~~ $1/\eps_0$~~~~~~&$1/\mu_0$~~~~~~~~1\\
Heaviside-Lorentz & ~~~~~~~ $1$~~~~~~~~& $1$~~~~~~~~~~~~~$1/c$\\
\hline
\end{tabular}
}
\end{center}
 \par From this Table one can see the possibility to introduce another set of constants with notation and presence in the different expressions closed to {\bf SI}:
 \beqa\label{newc}
 \tilde{\eps_0}\,=\,\frac{1}{\al},\;\;\tilde{\mu_0}\,=\,\frac{1}{\la},\;\;c_0\,=\,\frac{1}{\gam}\ .
 \eeqa

 With the notation \eref{newc}, the transcription of Table 1. is 
 
 \par 
\vspace{0.5cm}
\begin{center}
\par \scriptsize{Table 2. The $\tilde{\eps_0},\,\tilde{\mu_0},\,c_0$-system
\par
\begin{tabular}{@{}l@{}l @{}l}
\hline
System &~~~~~~~  $\tilde{\eps_0}~~~~~$ & $\tilde{\mu_0}$ ~~~~~~~$c_0$\\
\hline
Gauss &~~~~~~~ $1/4\pi$~~~~~~~~&$4\pi$~~~~~~~ $c$ \\
SI~~~ &~~~~~~~ $\eps_0$~~~~~~&$\mu_0$~~~~~~~~1\\
Heaviside-Lorentz & ~~~~~~~ $1$~~~~~~~~& $1$~~~~~~~~~~$c$\\
\hline
\end{tabular}
}
\end{center}

The Maxwell's equations with the new constants \eref{newc} are written as
\beqa\label{M-new}
\nablav\cdot\Ev\,=\,\frac{1}{\tilde{\eps_0}}\,\rho\ ,\nonumber\\
\nablav\cdot\Bv\,=\,0\ ,\nonumber\\
\nablav\times\Bv\,=\,\frac{\tilde{\mu_0}}{c_0}\left(\jv\,+\,\tilde{\eps_0}\,\frac{\d\Ev}{\d t}\right)\ ,\nonumber\\
\nablav\times\Ev\,=\,-\frac{1}{c_0}\,\frac{\d\Bv}{\d t}\ .
\eeqa
 \par The notation  \eref{newc} is used in  \cite{cvnat}. The motivation for such notation is a pedagogical one.  One of the first textbooks using such a system of notation is, by our knowledge, Ref. \cite{Novojilov}. In this textbook one uses the notation $\eps_0$ and $\mu_0$ instead of the notation \eref{newc}; these constants are introduced from the very beginning, as to account for the different demensions and units of measurements of electromagnetic quantities. This notation is used consistently in \cite{cvub} and \cite{cvedp} and is extended also to the relativistic covariant formulation of electrodynamics. However, this is only history.
 
  \par We resume  the basic expressions which are different from those written in {\bf SI} notation, the differences being reprezented by the presence of  $c_0$. In the following the notation $\eps_0,\,\mu_0$ is used instead of $\tilde{\eps_0},\,\tilde{\mu_0}$.
 \par They are: the densities of energy current and of linear momentum
 \beqan
 \bbox{S}\,=\,\frac{c_0}{\mu_0}\,\Ev\times\Bv\ ,\;\; \bbox{g}\,=\,\frac{\eps_0}{c_0}\,\Ev\times\Bv\ ,
 \eeqan
 the Lorentz force:
 \beqan
 \bbox{F}\,=\,q\Ev\,+\,\frac{1}{c_0}\,q\,\bbox{v}\times\Bv\ ,
 \eeqan
 the expression of the electric field in terms of potentials:
 \beqan
 \Ev\,=\,-\nablav\,\Phi\,-\,\frac{1}{c_0}\frac{\d\Av}{\d t}\ ,
 \eeqan
 the gauge transformations and Lorenz condition:
 \beqan
 \Av\,\longrightarrow\,\Av\,+\,\nablav\psi,\;\;\Phi\,\longrightarrow\,\Phi\,-\,\frac{1}{c_0}\frac{\d\psi}{\d t},\;\;\;
 \nablav\cdot\Av\,+\,\frac{\eps_0\mu_0}{c_0}\,\frac{\d\Phi}{\d t}\,=\,0\ ,
 \eeqan
 the retardation solution for the vector potential:
 \beqan
 \Av(\rv,\,t)\,=\,\frac{\mu_0}{4\pi c_0}\int\frac{[\jv]}{R}\rmd^3x'\ ,
 \eeqan
 with the corresponding consequences when writing the fields $\Ev$ and $\Bv$.
  \par In the relativistic covariant formulation of electrodynamics we have:
 the 4-potential
 \beqan
 \big(A^\mu\big)\,=\,\big(\frac{c_0}{c}\,\Phi,\,\Av\big)\ ,
 \eeqan
 and the field tensor
 \beqan
 \big(F^{\mu\nu}\big)\,=\,
 \left( 
\begin{array}{c|c}
          0     &   -\frac{c_0}{c}\,\Ev \\
 \hline \\
      \frac{c_0}{c}\,\Ev  & -\eps_{ijk}\,B_k                                 
\end{array} 
\right)\ .
\eeqan

From the above expression one can obtain all the expressions used in electromagnetic theory. We give in the present note only the transformation equations for the fields $\Ev$ and $\Bv$:
\beqan
\Ev'\,&=&\,\Gamma\,\Ev\,+\,\frac{1-\Gamma}{v^2}\,(\bbox{v}\cdot\Ev)\,\bbox{v}\,+\,\frac{1}{c_0}\,\Gamma\,(\bbox{v}\times\Bv)\ ,\\
\Bv'\,&=&\,\Gamma\,\Bv\,+\,\frac{1-\Gamma}{v^2}\,(\bbox{v}\cdot\Bv)\,\bbox{v}\,-\,\frac{c_0}{c^2}\,\Gamma\,(\bbox{v}\times\Ev)\ ,
\eeqan
where $\Gamma\,=\,1/\sqrt{1-v^2/c^2}$.
\par The 4-dimensional form of Maxwell's equations is written as
\beqa\label{4-dim}
\d_\nu\,F^{\nu\mu}\,=\,\frac{\mu_0}{c_0}\,J^\mu\ ,\nonumber\\
\d_\nu^{\;\;} ~^*F^{\nu\mu}\,=\,0\ ,  
\eeqa
where 
\beqan  (J^\mu)\,=\,(c\rho,\;\jv),\;\;\;~^*F^{\mu\nu}\,=\,\frac{1}{2}\,\eps^{\mu\nu\la\rho}\,F_{\la\rho}\ .
\eeqan
\par Even the quantum electrodynamics can be formulated in an independent of units notation. For electromagnetic field no additional  
rules are necessary and the Dirac equation for the electron can be writen as \cite{cv-04}
\beqan
\big(\rmi \hbar\,\hat{\d}\,+\,\frac{e}{c_0}\,\hat{A}\,-\,m_ec\big)\,\psi\,=\,0\ ,
\eeqan
$-e$ and $m_e$  being the electric charge and the mass of the electron. 

 \par In the present note we want to plead for using a notation system in the spirit or as close as possible to {\bf SI}. The pedagogical motivation is obvious. Such a formulation has also the advantage that one can simplify the life of a student  who does not consent the general manner of notation. From the very beginning, one can state the possibility of working only in {\bf SI}  units if setting everywhere  $c_0\,=\,1$. The student can have the freedom to choose the latter option without any repercussion from the professor, of course.

 \vspace{0.5cm}
\par 
\end{document}